\input{aipcheck}

\documentclass[
    ,final            
  ]
  {aipproc}

\layoutstyle{6x9}

\begin{document}

\title{LISA Observations of Supermassive Black Hole Growth}


\author{Miroslav Micic, Kelly Holley-Bockelmann and Steinn Sigurdsson }{
  address={Pennsylvania State University, 525 Davey Lab, University Park, 16802}
}



\begin{abstract}
Based on a high resolution cosmological n-body simulation, we 
track the hierarchical growth of black holes in galaxy clusters from $z=20$
to $z=0$. We present a census of black holes as 
function of redshift and 
will determine their mass assembly history under a variety of assumptions
regarding the importance of gas accretion in black hole growth, from early
supercritical Eddington accretion to gas-poor hierarchical assembly.
Following a galaxy merger, black holes are expected to 
form, inspiral and merge after strongly radiating energy via gravitational
waves. For each binary black hole inspiral and merger, we determine the 
expected gravitational wave signal for the Laser 
Interferometer Space Antenna (LISA), and calculate the LISA event 
rate as a function of time. We will calculate the black hole mass
assembly history for several black hole growth scenerios, so that we can
explore tests to characterize each model observationally.
In particular, we will study how well LISA observations 
will be able to distinguish between these very different assembly scenarios.
\end{abstract}

\maketitle


\section{MOTIVATION}
In current theories of structure formation, the first generation of star
formation seeds intermediate mass black holes deep within the potential of
forming galaxies (Madau $\&$ Rees 2001, Abel et al. 2002, Heger et al. 2003). 
Mass of the structures that can host first massive black holes is limited 
by various feedback processes. From Population III supernova studies, 
the formation of first stars stops at z$\sim$12 (Wise $\&$ Abel 2005). 
These black holes can grow in mass by accreting gas, and as the event 
horizon grows in size, stars and stellar remnants. Later, galaxy 
mergers provide an impetus for these growing black holes to meet and form 
a bound system (Volonteri et al. 2003, Islam et al. 2003). During a galaxy 
merger, each black hole sinks to the center of the new galaxy potential due to 
dynamical friction, and eventually becomes bound as a binary.
Dynamical friction then continues to shrink the orbit until the binary is hard
(i.e., the separation between each black hole, $a_{\rm BBH}$, is such that 
the system tends to lose energy during stellar encounters)(Madau $\&$ Rees 2001, 
Holley-Bockelmann et al. 2001, Vine $\&$ Sigurdsson 1998). Thereafter, 
further decay is mediated by 3-body 
scattering with the ambient stellar background until the binary becomes so 
close that the orbit can lose energy via gravitational radiation. In studies 
of a static, spherical potential, it may be difficult for stellar encounters 
alone to cause the binary to transition between the 3-body scattering phase and
the gravitational radiation regime; however in gas-rich or non-axisymmetric 
systems the binary hardens efficiently into one that emits copious 
gravitational radiation. Thereafter, it presumably coalesces. Depending on the 
black hole mass, the final stages of coalescence emit so much gravitational 
radiation (Thorne 1995, Cornish $\&$ Levin 2002, Vecchio et al. 2004) that they 
are extremely likely to be 
detected by the Laser Interferometer Space Antenna (LISA), a planned 
NASA mission to detect gravitational waves, set to launch in the next decade.
Current estimates of the total gravitational wave signal from the cosmological 
growth and merger lifecycle)of black holes have been made using semianalytic
models of merger trees, with analytic prescriptions for the black hole merger 
timescales within the analytic halos and models for gas accretion. In this
study, we use high resolution N-body simulations to track the seed black holes
themselves as they sink within a galaxy halo. We will then use this to
more accurately model different black hole growth scenerios, which will yield 
different gravitational wave signatures. Ultimately, we will determine if
LISA observations can be used as a tool to discover how black holes
grow.


\begin{figure}
 \includegraphics[height=.3\textheight]{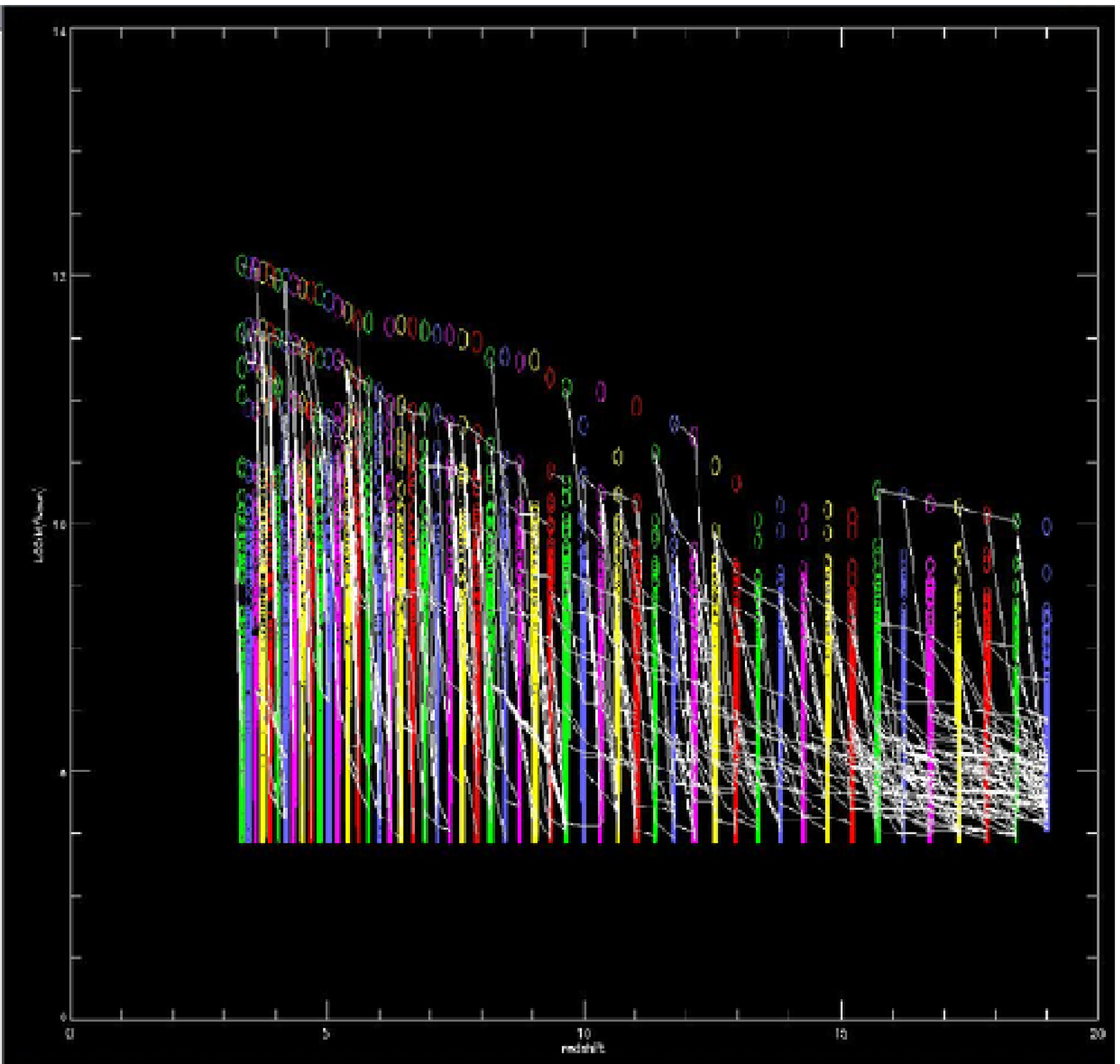} 
  \includegraphics[height=.3\textheight]{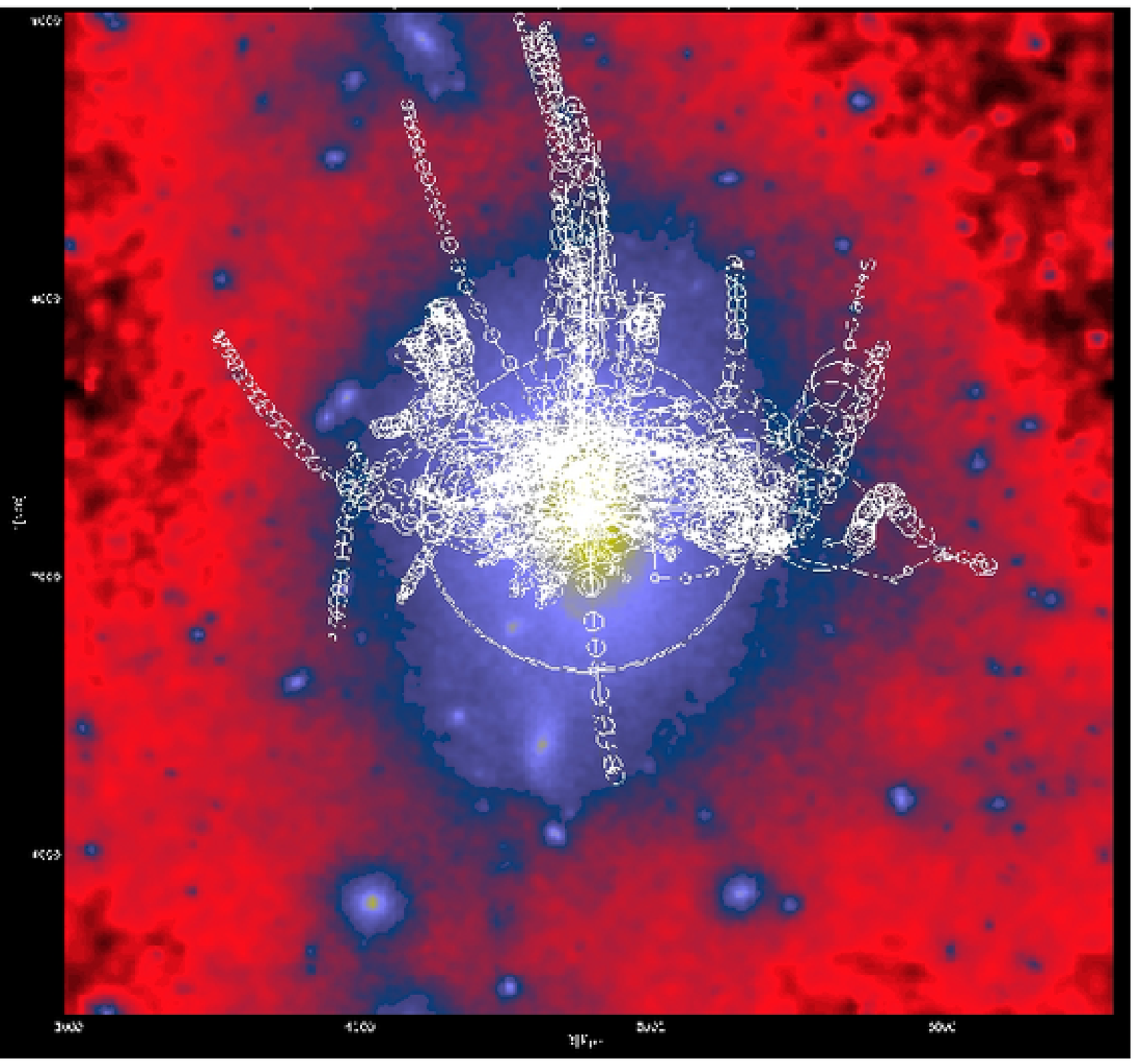}
 \caption{1a-left- Merger tree from $z=20$ to $z=3$. Different halos are 
represented with different colors. The sizes of the circles are set by the 
virial radius, which is related to $M_{\rm halo, virial}^{1/3}$. By our dry 
merger assumption, $M_{\rm BH1,}$ is also related to $M_{\rm halo, virial}$. 
Given this merger tree, we will test different models for black hole accretion. 
1b-right- 2D density projection of the simulation box with primary halo.
Overploted are paths of dark matter halos which grow in mass (circles represent scaled
virial radii) and sink toward primary. Only main branches of the merger tree are presented. }
\end{figure}

\section{Growing Supermassive Black Holes}
In our numerical simulations we use GADGET to evolve a comoving 14.3 Mpc$^3 $
section of a $\Lambda$CDM universe ($\Omega_M$=0.3, $\Omega_{\Lambda}$=0.7 
and h=0.7) from $z=40$ to $ z=0$. We refine a sphere of 2 Mpc in the box to 
simulate at higher resolution with
4.9$\times10^6$ high-resolution particles (softening length 2 kpc
comoving). The rest of the box has  2.0$\times10^6$ low-resolution
particles (softening length 4 kpc comoving).
The mass of each high resolution particle in this simulation is
8.85$\times10^5$M$_\odot$ and the mass of each low-resolution particle is
5.66$\times10^7$M$_\odot$. We use P-GroupFinder to define dark matter halos 
and to identify black holes as the most bounded particles in their host halos.

Only dark matter halos with mass M$_{vir}$$\geq$10$^8$M$_\odot$$[(1+z)/10]$$^{-1.5}$
can host first stars and the first stars stop forming at z$\sim$12. By tracking 
the positions of over 3800 seed black holes and their
host halos throughout the simulation (Fig. 1), we will model different black hole
merger scenerios and track the growth of the black hole mass per halo as
a function of time and environment. In particular, we will parameterize the 
degree of gas accretion involved with growing the black hole and the efficiency
of binary coalescence.

In this first attempt, however, we model the seed black hole merger timescale 
and the mass accretion history in the simplest possible way. We assume that
the hardening timescale is rapid, that no black holes are ejected, and that 
there is no gas accretion involved. This gives us a fiducial model with which
to compare more realistic black hole growth scenerios. 

\begin{figure}
 \includegraphics[height=.2\textheight]{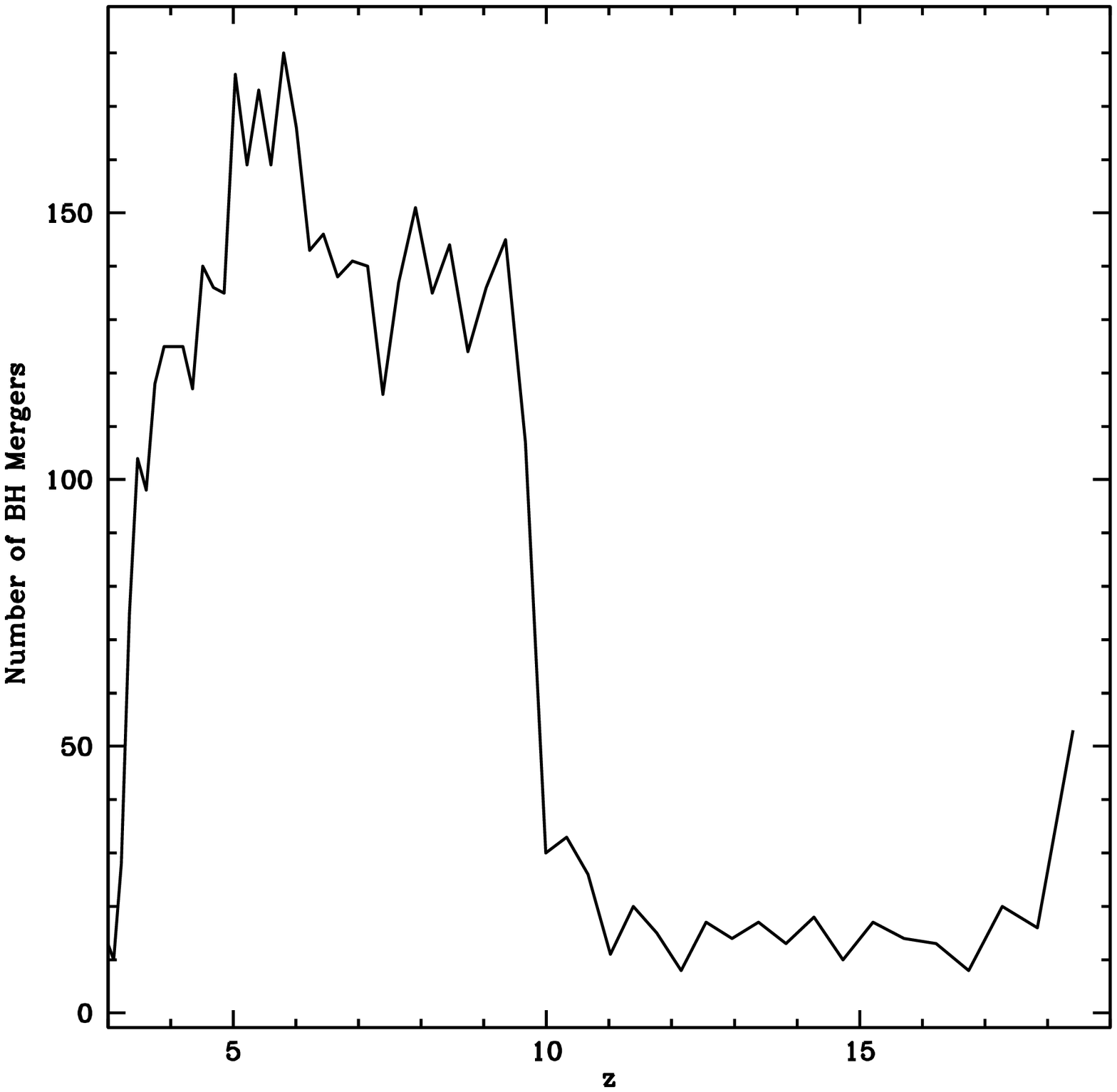} 
  \includegraphics[height=.2\textheight]{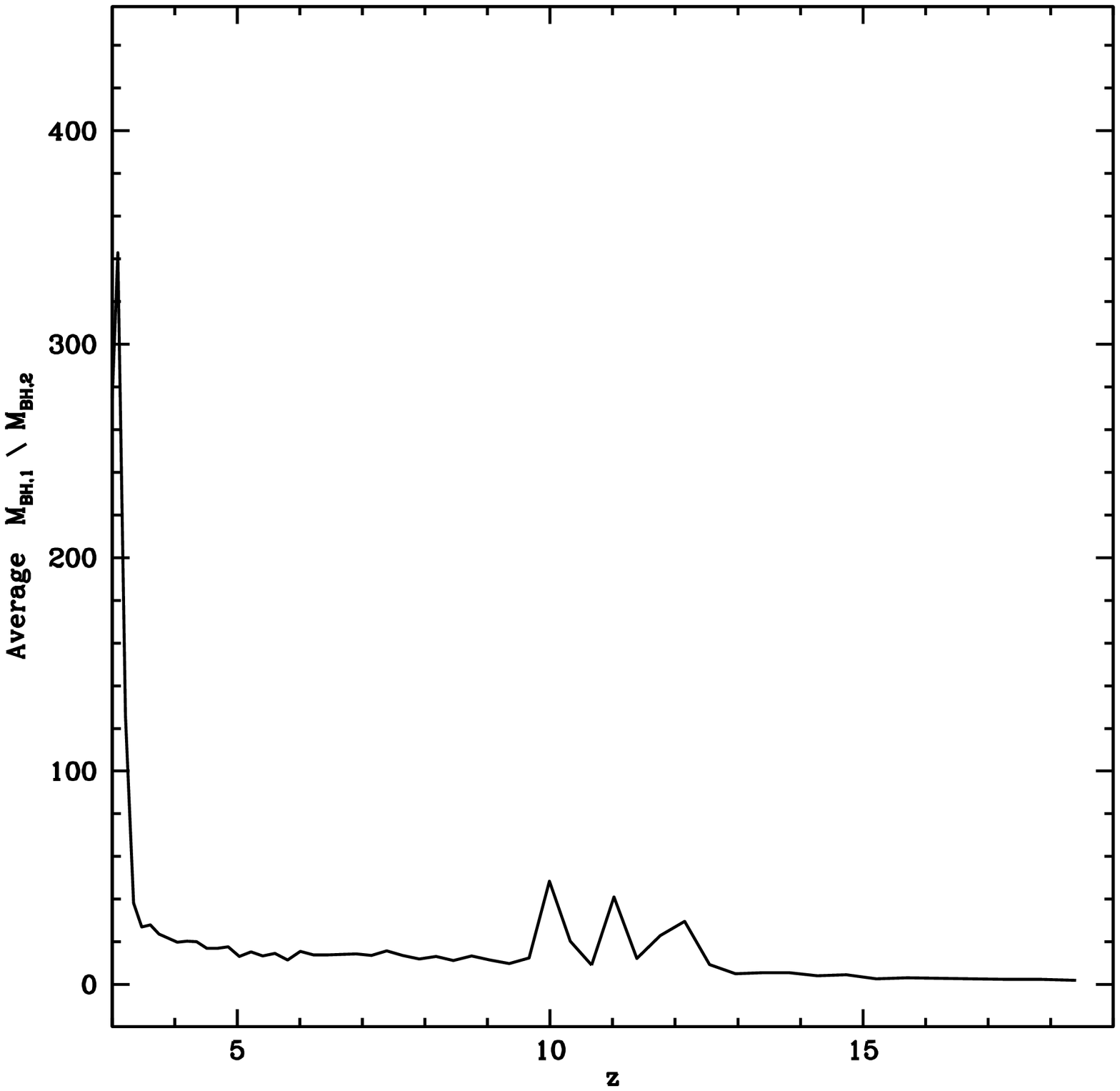}
 \includegraphics[height=.2\textheight]{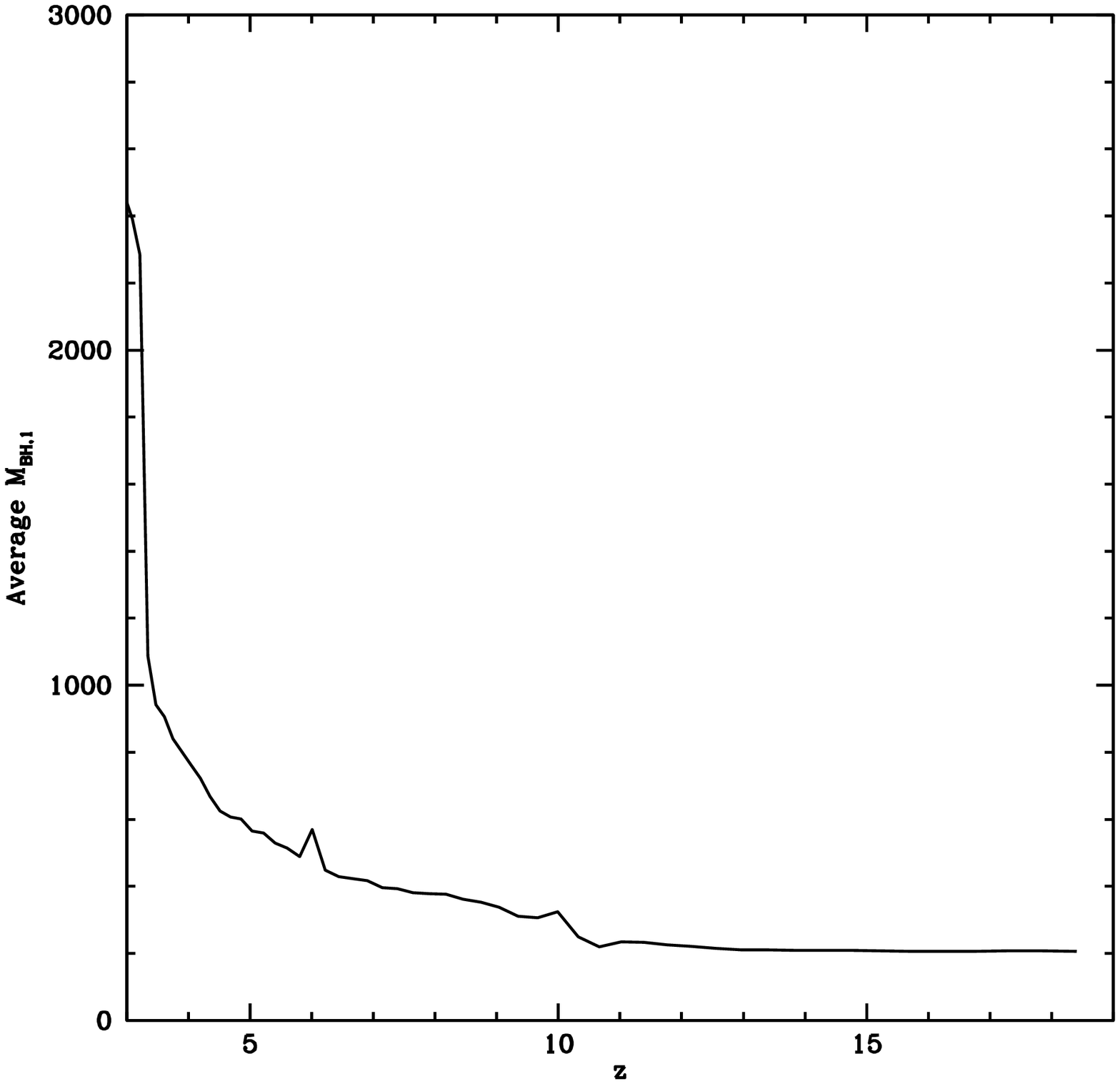} 
 \caption{2a-Left- Number of IMBHs mergers as a function of redshift. 2b-Middle- 
 Average ratio of merging black holes as a function of redshift. 2c-Right-  
Average of more massive components in the merger as a function 
of redshift.} 
\end{figure}

\section{Dry Merger And Mergers Plus Accretion} 

We start with 200M$_\odot$ initial black hole mass.
Given this simple growth scenerio, we find a large number of mergers at each 
redshift, with the largest number of mergers at $z\sim 6$. At later times, 
$100:1$ mergers are much more common, as $M_{\rm BH1}$ grows larger while 
also accreting smaller mass black holes from satellite halos (Fig. 2). At this stage we 
have analyzed the data in range 3$\leq$z$\leq$20. The intensity of mergers of
dark matter structures demands more detailed analysis, to be addressed in the
future work. Notice that dry mergers build massive black holes M$\geq$10$^6$M$_\odot$
by redshift z=3. We add accretion to the previously described growth of massive black holes
through mergers. For a sustained Eddington accretion of baryon matter the mass
growth rate is M$_{BH}$ / t$_{Sal}$, 
where t$_{Sal}$ is Salpeter time-scale, t$_{Sal}$$\sim$4$\times10^7$ yr from
the recent observations. Accretion is triggered by mergers of dark matter halos
and lasts t$_{Sal}$. During this time black holes doubles its mass. 

\begin{figure}
 \includegraphics[height=.2\textheight]{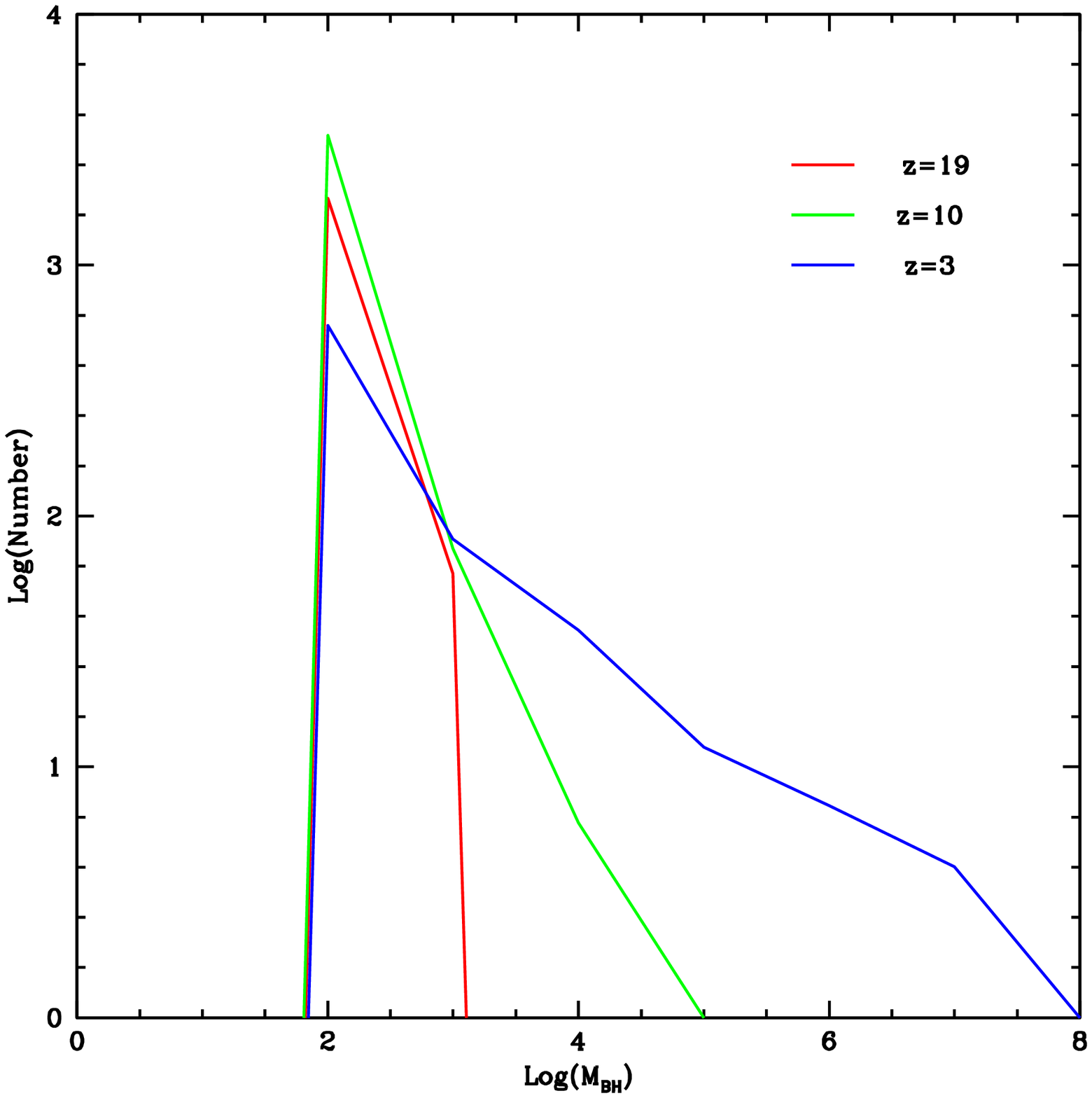} 
 \includegraphics[height=.2\textheight]{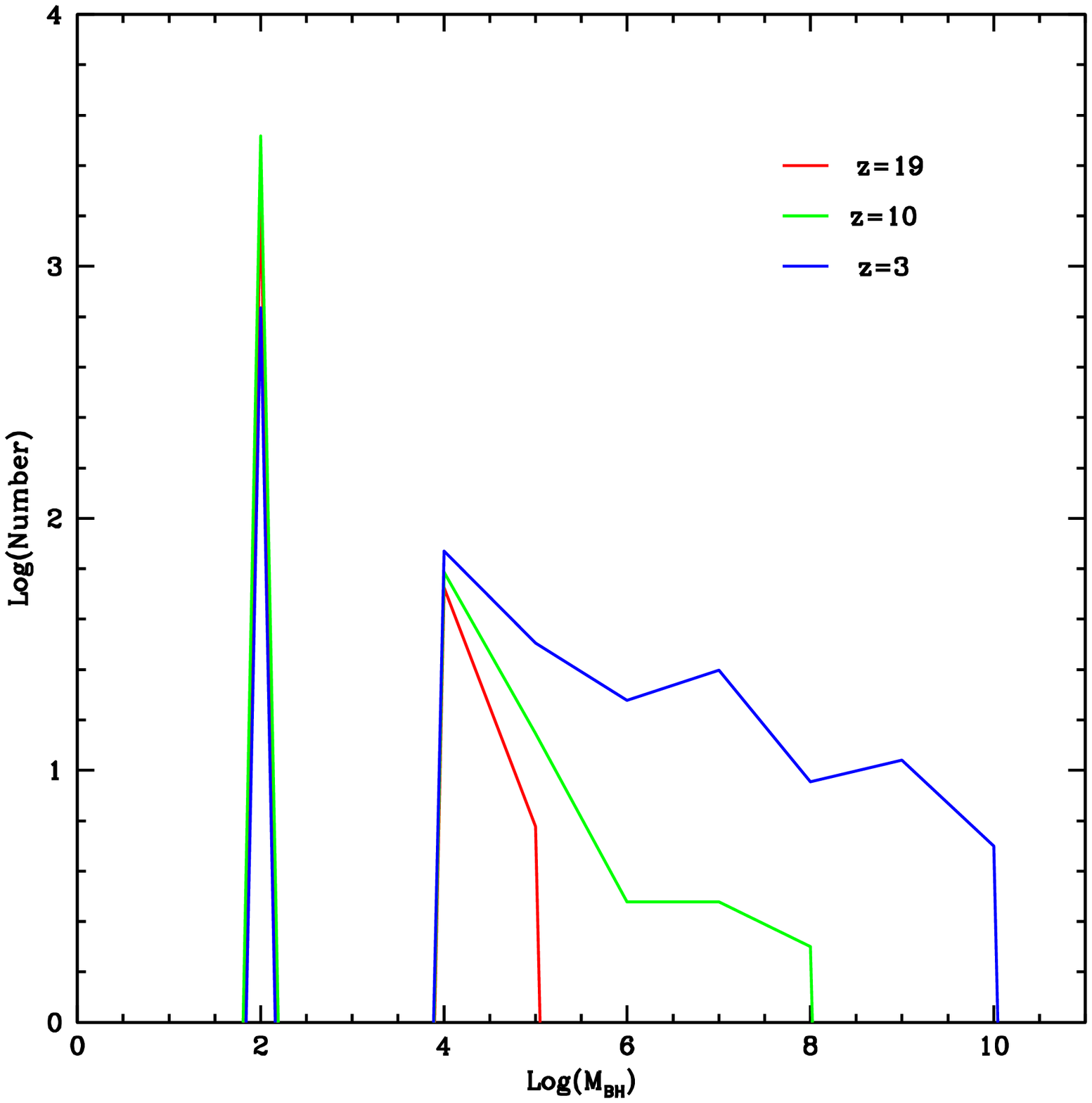} 
 \caption{3a-left- Mass function for merging black holes. Red - redshift z=19;
Green - redshift z=10;  Blue - redshift z=3. 3b-right- Mass function for massive 
black holes which grew through mergers and accretion. There is a number of 
supermassive black holes at redshift z=3. }
\end{figure}

\section{Extracting Gravitational Waves}

We assume the black hole binary only emits gravitational radiation
during the 'gravitational wave' phase, where the orbital separation
of the binary decays most rapidly due to gravitational wave emission.
In reality, this is a lower limit to the amount of gravitational radiation
emitted during a typical binary's evolution because gravitational radiation
is also produced during the hardening phase, as well as when the binary
coalesces and rings down. 

Under this assumption, the maximum rest frame frequency {\it f}$_r$, occurs
for a circular orbit at 3 Schwarzschild Radii:

\begin{equation}
{f_{max}}=\frac{c^{3}}{14.7{\pi}G}\frac{(M_{1}+M_{2})^{1/2}}{M_{1}^{3/2}}
\end{equation}

where M$_1$, and M$_2$ are the black hole masses, G is the gravitational
constant, and c is the speed of light. The minimum {\it f}$_r$ is:

\begin{equation}
{f_{min}}=\frac{1}{{\pi}}\frac{G{(M_{1}+M_{2})^{1/2}}}{a_{gw}^{3}}
\end{equation}

where a$_{gw}$ is the binary separation where gravitational radiation
dominates, expressed by:

\begin{equation}
{a_{gw}}=0.0014pc[\frac{M_{1}M_{2}(M_{1}+M_{2})}{10^{18.3}{M_\odot}^3}]^{1/4}{t_{g}}^{1/4}
\end{equation}

where t$_g$$^{1/4}$ is the coalescence timescale in Gyr. The change in the rest frame frequency is: 
%

\begin{equation}
{\dot{f}_r}=\frac{96}{5}{f_r}^{5/3}\frac{G^3M_{1}M_{2}(M_{1}+M_{2})}{c^5a^3}[\frac{G(M_{1}+M_{2})}{{\pi}^2}]^{-1/3}
\end{equation} 

So, for each merger at a comoving distance d(z), gravitational radiation
is emitted with a characteristic strain, h$_c${\it f}$_r$

\begin{equation}
{h_c(f_r)}=\frac{g(\epsilon)8{\pi}^{2/3}G^{5/3}{M_{chirp}}^{5/3}{f_r}^{5/3}}{3.16c^4d(z){\dot{f}_r}}
\end{equation}


\begin{figure}
 \includegraphics[height=.2\textheight]{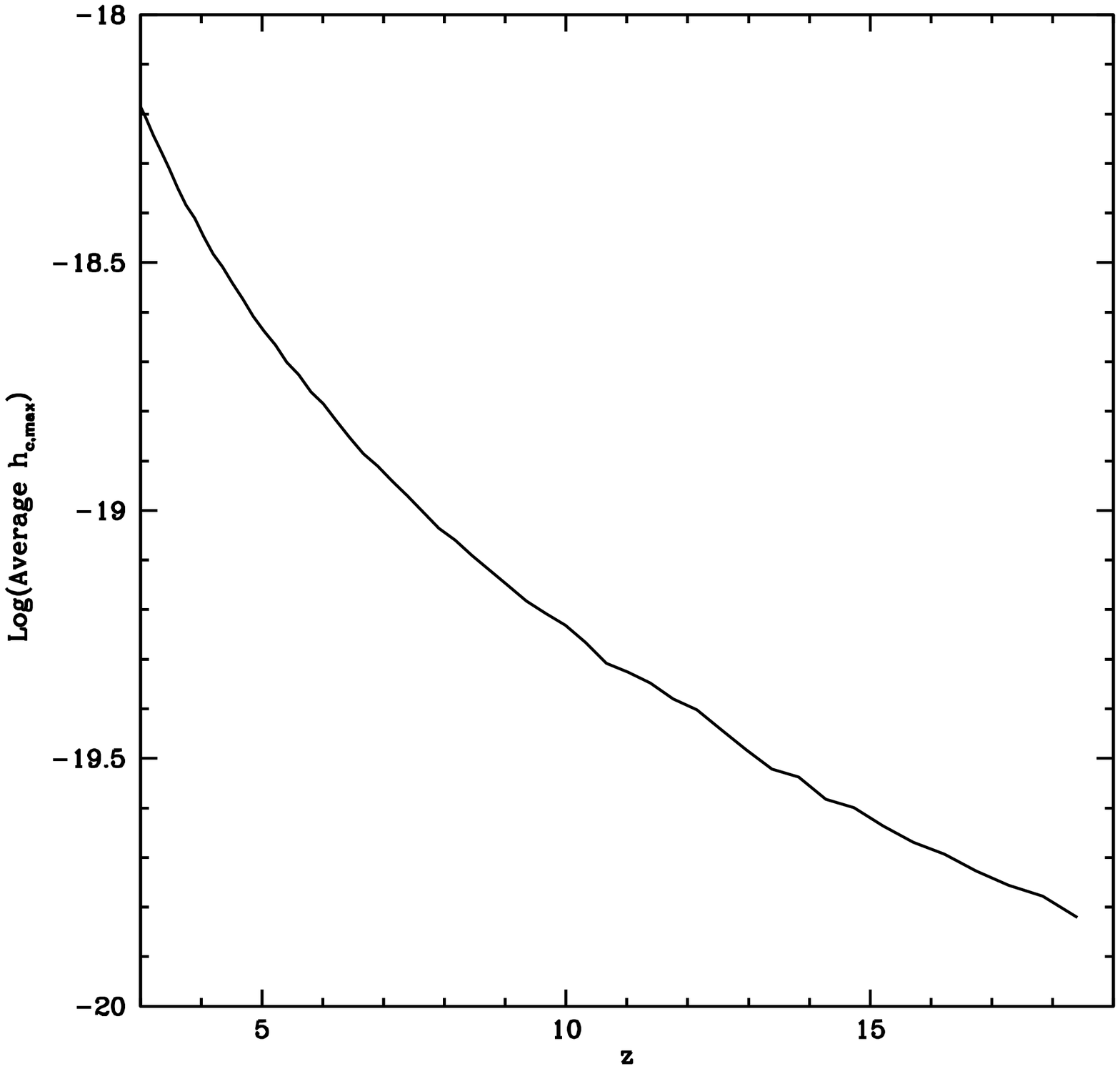} 
  \includegraphics[height=.2\textheight]{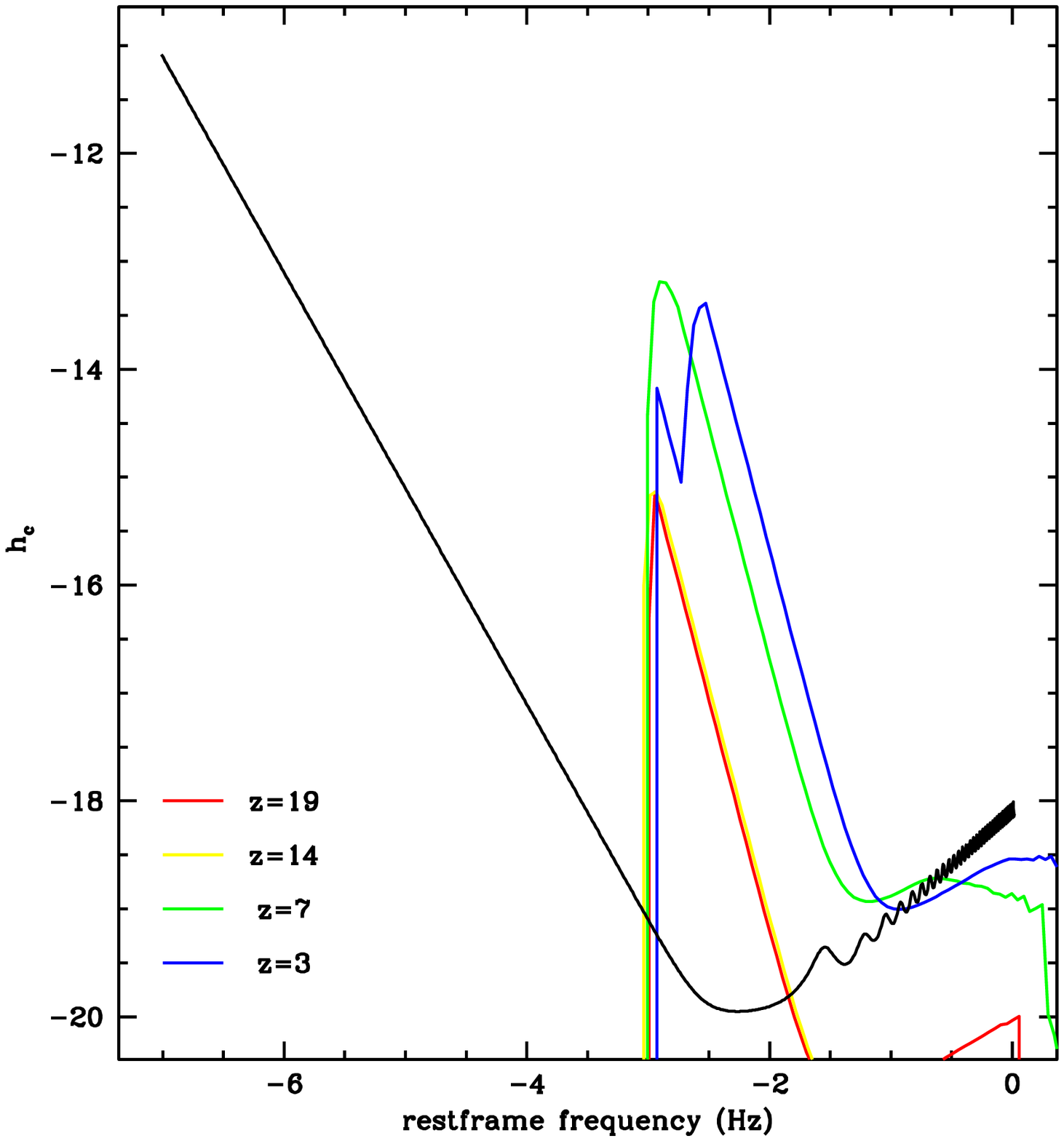}
\end{figure}
\begin{figure}
 \includegraphics[height=.2\textheight]{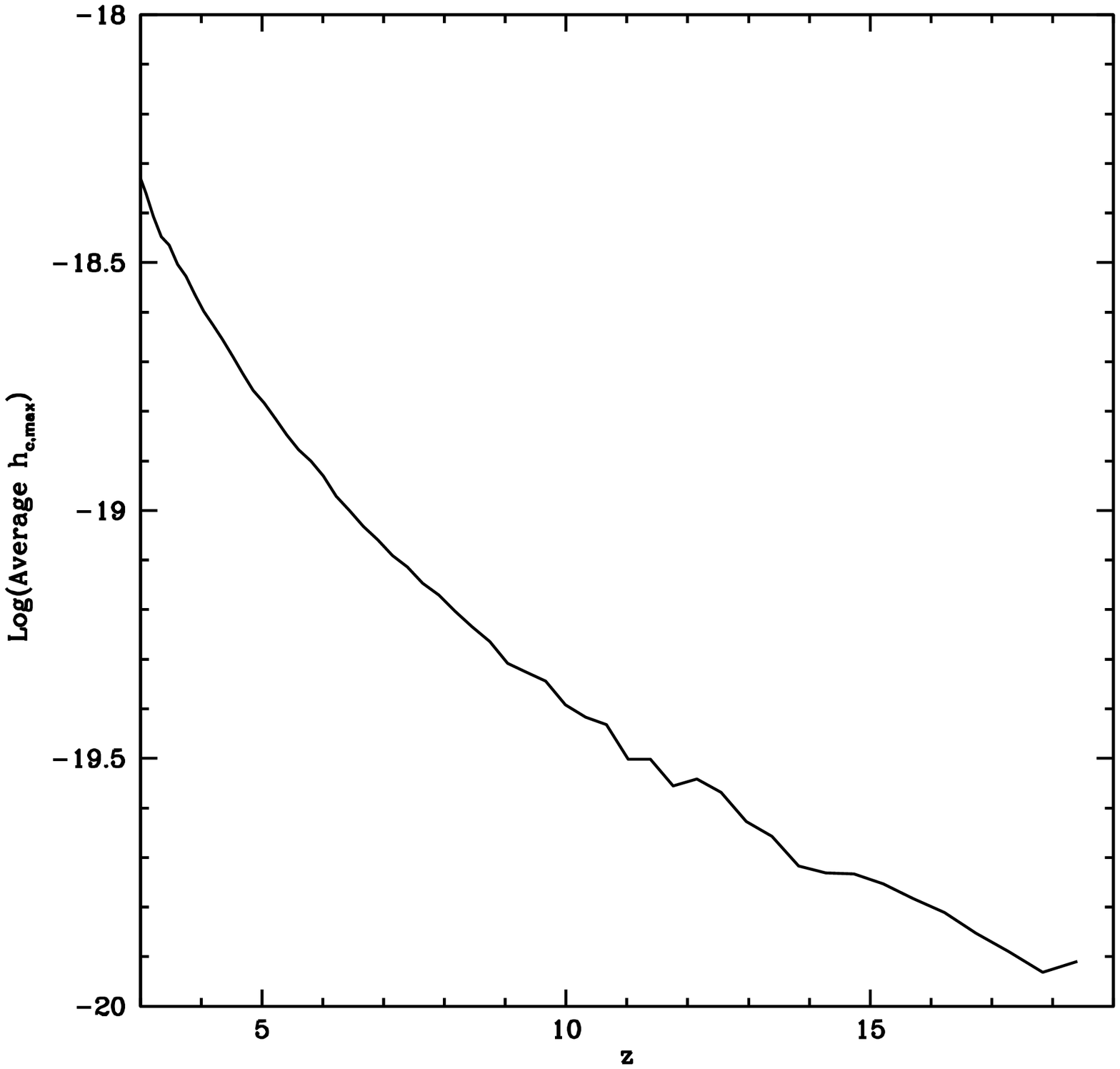} 
  \includegraphics[height=.2\textheight]{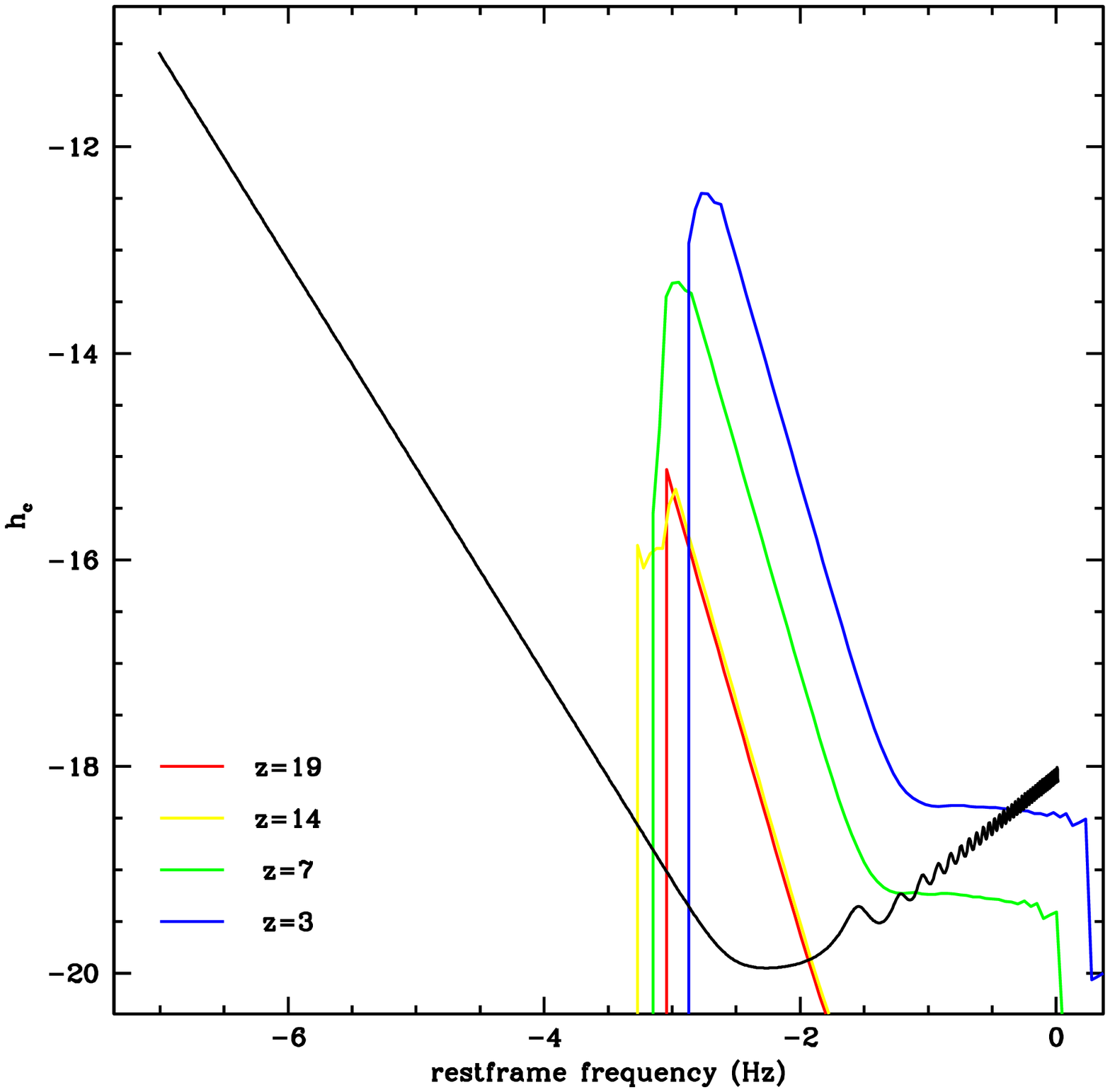}
 \caption{4a-up left- shows the average observed characteristic gravitational 
strain amplitude as a function of observed redshift. Distance is measured from the center
of the primary. Gravitational strain amplitude increases toward smaller redshifts 
as IMBHs sink into gravitational potential of the primary. This increase is rapid
since the ratio of IMBHs binaries increases too. 4b-right- LISA sensitivity curve 
(black) and the total gravitational strain amplitude at different redshifts. The 
increase in mass of merging black holes with redshift results in increasing total 
gravitational strain amplitude and shifting toward higher observed frequencies. 
4c and 4d on the bottom - gas accretion added.} 
\end{figure}


\begin{theacknowledgments}
We acknowledge the support of the Center for Gravitational Wave
Physics funded by the NSF under cooperative agreement PHY 01-14375,
NSF grants PHY 98-00973.
\end{theacknowledgments}



\bibliographystyle{aipproc}   

\bibliography{sample}

\IfFileExists{\jobname.bbl}{}
 {\typeout{}
  \typeout{******************************************}
  \typeout{** Please run "bibtex \jobname" to optain}
  \typeout{** the bibliography and then re-run LaTeX}
  \typeout{** twice to fix the references!}
  \typeout{******************************************}
  \typeout{}
 }
\end{document}